# Electron spin echo relaxation and envelope modulation

# of phosphorus shallow donors in silicon

Anna Ferretti and Marco Fanciulli*
*Laboratorio Nazionale MDM – CNR-INFM, via C. Olivetti 2, I-20041 Agrate Brianza, Italy*

Alessandro Ponti*
*Istituto di Scienze e Tecnologie Molecolari – CNR, via C. Golgi 19, I-20133, Milano, Italy*

Arthur Schweiger
*Department of Chemistry and Applied Biosciences, ETH Hönggerberg, CH-8093 Zürich, Switzerland*

* Electronic addresses: marco.fanciulli@mdm.infm.it, alessandro.ponti@istm.cnr.it

**Abstract**
Spins of single donor atoms are attractive candidates for large scale quantum information processing in silicon, since quantum computation can be realized through the manipulation of electron and/or nuclear spins. We here report on two-pulse electron spin echo experiments on phosphorus shallow donors in natural and $^{28}$Si-enriched silicon epilayers doped with $10^{16}$ cm$^{-3}$ P donors. The experiments address the spin-spin relaxation times and mechanisms and provide, through the electron spin echo envelope modulation (ESEEM) effect, information on the donor electron wave function. Experiments performed as a function of the pulse turning angle allowed us to measure the exponential relaxation and spectral diffusion times depurated by instantaneous diffusion. According to these results, isotopically purified samples are necessary to reduce the spectral diffusion contribution and the P shallow donors concentration plays a fundamental role in determining the intrinsic phase memory time. ESEEM peaks have been assigned to hyperfine-coupled silicon-29 nuclei at specific crystallographic positions on the basis of a spectral fit procedure including instrumental distortions.

# 1 Introduction

Spins of single donor atoms are attractive candidates for large scale quantum information processing (QIP) in silicon or silicon-germanium. In several solid state schemes, quantum computation is implemented through the manipulation of electron and/or nuclear spins of single donors ($^{31}$P) in silicon[1] or in SiGe heterostructures,[2,3] double donors ($^{126}$Te) in Si,[4] or electrostatically confined electron spins in quantum dots.[5] In the original proposal by Kane,[1] the qubits are the *nuclear* spins of $^{31}$P donors in silicon. However, low nuclear spin transition frequencies (MHz range) and slow relaxation result in a low clock rate. Orders of magnitude higher clock rates are possible if *electron* spins are used as qubits instead of nuclear spins.[2] Quantum computation also requires that qubits interact in a controlled way to perform two-qubit operations. In solid state schemes, spins usually interact through the exchange interaction that can be modulated by electrical gates. In bulk silicon, this requires that the donor atom spacing is about 15 - 20 nm so that the electron-electron exchange interaction strength is in the 0.1 meV range.

Fault-tolerant QIP is possible only if the qubit coherence time is long compared to a $\pi$ pulse.[6] This requirement can be satisfied in silicon at low temperatures (< 5 K).[7] The low concentration of magnetic nuclei ($^{29}$Si $I = 1/2$, nat. ab. = 4.67 %) makes silicon a better candidate than III-V compounds.[8] Several mechanisms may play an important role in the loss of electron spin coherence, depending on the qubit structure, so that the qubit coherence time must be addressed in conditions as close as possible to the real device. Electron paramagnetic resonance (EPR) is the best technique to investigate electron spin relaxation because it directly monitors the magnetization associated to the spins. Early continuous-wave EPR measurements revealed that the resonance lines are inhomogeneously broadened due to unresolved super-hyperfine interaction, and that they are affected by the Si isotopical purity.[9] Because of the line inhomogeneity, the transverse relaxation time $T_2$, representing the electron spin coherence time scale, cannot be accurately determined by continuous-wave techniques. As this parameter is critical for QIP, it must be determined with more accurate techniques such as pulse EPR. Early pulse EPR experiments[10] showed that also the homogeneous line width depends on the concentration of $^{29}$Si: the phase memory time $T_m$ at 1.4 K resulted ~200 μsec for natural Si, and ~520 μsec for isotopically purified Si. Later electron spin echo experiments[11] performed as a function of the doping level gave comparable results and confirmed the non-single exponential decay of the echo, a result attributed to spectral diffusion.[12,13] In addition to relaxation, decoherence is also induced by the non-diagonal terms in the hyperfine interaction which entangle the donor electron spin and the $^{29}$Si nuclei.[8,13]

We report on two-pulse electron spin echo experiments on P shallow donors in natural and isotopically pure $^{28}$Si. The experiments address the spin-spin relaxation times and mechanisms and provides, through the electron spin echo envelope modulation (ESEEM) effect, information on the donor electron wave function. The latter has been extensively investigated in Si by the electron-nuclear double resonance (ENDOR) technique.[9,14] Recently, electron spin phase relaxation of P donors and ESEEM results have been reported in natural and $^{29}$Si depleted,[7,15] and in $^{29}$Si enriched[16] silicon crystals. Since the P and $^{29}$Si concentrations have a significant influence on the electron spin relaxation time, we studied samples with concentrations suitable for the implementation of a silicon-based quantum computer.

## 2 Experimental

The natural silicon (Si-nat) and the $^{28}$Si-enriched silicon (Si-28) epilayers [50 µm thick on a Si(100) p-type highly resistive substrate] were produced by ISONICS. The concentration of $^{29}$Si in the Si-28 epilayer is below 0.1%. The phosphorus concentration was found to be [P] = $4.1 \times 10^{16}$ cm$^{-3}$ in Si-nat and $3.3 \times 10^{16}$ cm$^{-3}$ in Si-28 by Hall-effect measurements.[7] These concentrations correspond to mean P-P distances $<r> \approx (2\pi N)^{-1/3}$ of 16 and 17 nm, respectively, as required for the implementation of a quantum computer.

Pulse EPR measurements were performed on a Bruker E580 spectrometer at 10 and 7 K with the (100) crystallographic axis of the substrate oriented along the static magnetic field. The electron spin echo decay was measured by the two-pulse sequence $\beta/2-\tau-\beta-\tau-$echo with $\tau_{min}$ = 350 ns and dwell time $\Delta\tau$ = 1 µs. The decay was recorded for several pulse turning angles $\beta$. The pulse lengths were 8 and 16 ns, so that one of the hyperfine lines of the EPR spectrum is uniformly excited. The microwave power was calibrated by searching for the maximum echo amplitude, corresponding to $\beta = \pi$. Smaller turning angles were obtained by decreasing the microwave power of both pulses.

Two-pulse ESEEM was recorded using the sequence $\pi/2-\tau-\pi-\tau-$echo with $\tau_{min}$ = 270 ns and a dwell time $\Delta\tau$ = 50 ns. The echo decay was eliminated by subtracting a fitted bi-exponential decay. Before FT, the time trace was apodized by a Kaiser $2\pi$ window,[17] a general-purpose apodizing function with high wiggle suppression that yields a good ESEEM spectrum from 0 to 10 MHz. To get a well-resolved ESEEM spectrum near $2\nu(^{29}$Si$)$ = 5.84 MHz, resolution enhancement was achieved by multiplying the time trace by a growing exponential (time constant 15 µs).[18] In both cases zero-filling up to 4096 points was carried out.

The fitting of the ESEEM spectra is based on the hyperfine matrices given in the literature.[14,19] ESEEM time traces were computed by the standard formulas for an $S$ = 1/2 electron spin (with isotropic Zeeman interaction) coupled to several $I$ = 1/2 nuclear spins as a function of the orientation of the static magnetic field in the crystal reference frame, represented by the polar angles $\theta$ and $\phi$. They were truncated to reproduce deadtime distortions and artifacts and then subjected to the same manipulation and Fourier transformed as the experimental data, and compared to the experimental ESEEM spectra using least squares optimization by the robust Nelder-Mead Simplex algorithm.[20] Parameter errors were estimated by the bootstrap method[21] with 101 samples.

## 3  Results and discussion

In this section we report and discuss the electron spin echo decay and the electron spin echo modulation data.

### 3.1  Electron spin echo decay

In a previous paper[7] it was demonstrated that the electron spin echo decay of phosphorus shallow donors in silicon is strongly influenced by spectral diffusion caused by the silicon nuclear spins. Indeed, echo decays can not be fitted by simply using a mono-exponential decay. The electron spin echo decay due to nuclear-spin induced spectral diffusion can take different analytical forms depending on the properties of the studied system and on the time scale of the experiment. Early theory[12] of spectral diffusion for the presently investigated time scale leads to echo decay of the form $\exp(-\tau^2)$ or $\exp(-\tau^3)$ for Lorentzian or Gaussian diffusion kernels, respectively. However, a detailed model[22] of nuclear-spin induced spectral diffusion in the electron spin echo decay of phosphorus donors in silicon shows that the contribution of spectral diffusion to the echo decay has the $\exp(-\tau^3)$ dependence for $\tau$ values well into the millisecond range, when the echo observed in our samples has largely decayed into background noise. Therefore, we use the augmented model[11]

$$I_e(\tau) = I_e(0) \exp[-2\tau/T_m - (2\tau/T_{SD})^3] \tag{1}$$

where $I_e$ is the echo intensity, $T_{SD}$ is the spectral diffusion characteristic time and $T_m$ is the exponential decay time constant. There may be several contributions to $T_m$, such as lifetime broadening ($2T_1$), flip-flop spin-spin relaxation ($T_2$), and in general all relaxation processes, also involving non-excited electron and nuclear spins, that give rise to an exponential decay. In addition, there is a contribution known as instantaneous diffusion ($T_{ID}$), which is not related to intrinsic spin relaxation, but is due to the modulation of dipolar couplings between electron spins caused by the microwave pulses. Since we shall depurate $T_m$ from instantaneous diffusion contributions, we define $T_{m0}$ as an intrinsic phase memory time

$$\frac{1}{T_m} = \frac{1}{T_{m0}} + \frac{1}{T_{ID}} \tag{2}$$

which can be operationally defined as the phase memory time in the limit of vanishing microwave pulse strength. Under several assumptions, $T_{ID}$ is given by[23]

$$\frac{1}{T_{ID}} = \frac{\pi}{9\sqrt{3}} C \frac{\mu g^2 \beta_e^2}{\hbar} \sin^2\left(\frac{\beta}{2}\right) \tag{3}$$

where $C$ is the concentration of the excited electron spins ($C = [P]/2$ in the present case since only one of the two hyperfine lines is excited), $\mu$ is the permeability of crystalline silicon, $g$ is the $g$-factor of the donor electron, $\beta_e$ is the Bohr magneton, and $\beta$ is the turning angle of the second microwave pulse. If the assumption of uniform excitation is relaxed, the expression for $T_{ID}$ is somewhat more involved[24]

$$\frac{1}{T_{ID}} = \frac{\pi}{9\sqrt{3}} C \frac{\mu g^2 \beta_e^2}{\hbar} \left\langle \sin^2\left(\frac{\beta_k}{2}\right) \right\rangle_f \tag{4}$$

where $\langle\ \rangle_f$ denotes averaging over the EPR line-shape. The turning angle $\beta_k$ is given by

$$\cos\beta_k = 1 - \frac{2\omega_1^2}{\Omega_k^2 + \omega_1^2} \sin^2\left(\sqrt{\Omega_k^2 + \omega_1^2}\, \frac{t_p}{2}\right) \tag{5}$$

where $t_p$ and $\omega_1$ are the length and strength (in angular frequency units) of the second microwave pulse ($\beta = \omega_1 t_p$) and $\Omega_k$ is the offset of the $k$-th spin contributing to the inhomogeneous EPR line.

In order to separate the contribution of instantaneous diffusion from $T_m$, primary echo decays were measured as a function of the pulse turning angle β. The echo decays recorded for Si-nat and Si-28 at 10 K or 7 K with pulse angles β = 180° or 90° are shown in Figure 1. Inspection of the decays observed for sample Si-nat at 10 K shows two distinct time intervals: the first one from the beginning of the echo recording (τ ≅ 0.34 μs) up to 2τ ≅ 25 μs, characterized by a very fast decay, and a second one from 2τ ≅ 25 μs to the complete decay of the electron spin echo, which could not be fitted to a single exponential. The initial fast decay is exponential with decay time ≅5 μs and reduces the initially observed global echo amplitude by 40% within 5 μs. It has been attributed to the modulation of the isotropic hyperfine interaction by the nuclear spins.[25] Even if the initial fast decay was not observed at 7 K, the first 25 μs of *all* time traces have been removed before the fitting process to avoid biasing. The stability of the optimized parameters was checked by fitting the time traces at 7 K after elimination of only the first 5 μs of the trace.

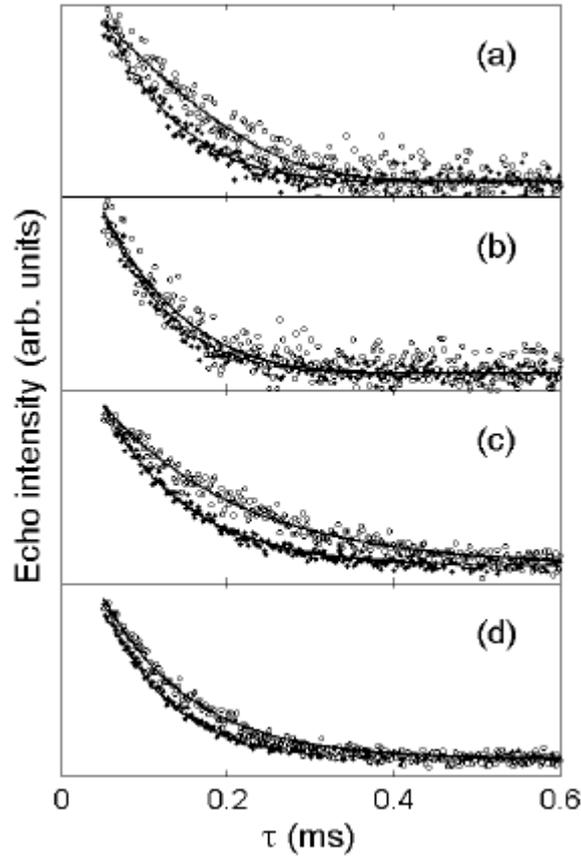

**FIG 1.** Effect of temperature and microwave pulse turning angle on the electron spin echo decay of Si-nat and Si-28 samples observed by the β/2–τ–β–τ–echo pulse sequence. a) Si-nat at 7 K; b) Si-nat at 10 K; c) Si-28 at 7 K; d) Si-28 at 10 K. Open circles: experimental data for β = 90°; solid circles: experimental data for β = 180°; lines: fitted curves. The decays for β = 90 and 180° have been arbitrarily rescaled for the sake of clarity. The β = 90° trace actually is about 2√2 ≅ 2.8 times less intense than shown.

First, we have individually fitted the experimental decays to Eq. (1) and obtained $T_m$ and $T_{SD}$ values for each curve. As the turning angle decreases, $T_m$ increases as expected and $T_{SD}$ shows a somewhat irregular variation. The usually employed procedure[15,24] would now prescribe a linear regression for $T_m(\beta)$ versus $\sin^2(\beta/2)$ to get $T_{ID}$ and $T_{m0}$ at each temperature $T$. The two-step nature of these procedure is, however, not necessary and could introduce errors because any mismatch between the experimental decays and the model described by Eq. (1) is arbitrarily assigned to a $T_{SD}$ variation with $\beta$. We therefore used a different approach, where the echo decay is considered as a function of the three variables $\tau$, $\beta$, and $T$, and the experimental decays are simultaneously fitted to a model equation obtained merging Eqs. (1), (2), and (3) [or (4) for the non-uniform excitation case]. In the former simpler case, the model equation is

$$I_e(\tau,\beta,T) = I_e(0,\beta,T)\exp\left\{-2\tau\left[\frac{1}{T_{m0}(T)} + C_1\sin^2\left(\frac{\beta}{2}\right)\right] - \left(\frac{2\tau}{T_{SD}(T)}\right)^3\right\} \quad (6)$$

where the dependences on $\beta$ and $T$ are put in evidence and $C_1$ is implicitly defined through Eq. (3). Of course, in Eq. (6) the relaxation times $T_{ID}$ and $T_{m0}$ depend on temperature, whereas $C_1$ is considered to be temperature-independent, since its variation in the investigated temperature range is negligible. This novel fitting procedure directly yields, as $\beta$-independent parameters, the spectral diffusion time $T_{SD}$, the intrinsic phase memory time $T_{m0}$, and the concentration of excited spins $C$ (via $C_1$).

**Table I**. Results of the simultaneous fitting of two-pulse electron spin echo decays of phosphorus shallow donors in silicon at different temperatures and second-pulse turning angles to the model Eqs. (6) and (7). The intrinsic, $T_{m0}$, and the conventional, $T_m(\pi)$, phase memory times, as well as the spectral diffusion time $T_{SD}$ are reported along with the phosphorus concentration $[P] = 2C$. The latter is compared with the value from Hall-effect measurements.

| Sample | $T$ (K) | $T_{m0}$ (ms) | $T_m(\pi)$ (ms) | $[P] = 2 C$ ($10^{16}$ cm$^{-3}$) | $[P]$ from Hall effect ($10^{16}$ cm$^{-3}$) | $T_{SD}$ (ms) |
|---|---|---|---|---|---|---|
| Si-nat | 10 | 0.17 ± 0.04 | 0.081 ± 0.004 | 1.5 ± 0.2 | 4.1 | 0.29 ± 0.03 |
|  | 7 | 0.22 ± 0.03 | 0.092 ± 0.005 |  |  | 0.33 ± 0.02 |
| Si-28 | 10 | 0.19 ± 0.03 | 0.074 ± 0.005 | 2.0 ± 0.1 | 3.3 | 0.4 ± 0.1 |
|  | 7 | 0.9 ± 0.2 | 0.109 ± 0.006 |  |  | 0.83 ± 0.04 |
| Si-28 [b] (w/out SD) | 10 | 0.15 ± 0.05 | 0.072 ± 0.002 | 1.7 ± 0.1 | 3.3 | — |
|  | 7 | 0.49 ± 0.02 | 0.109 ± 0.004 |  |  | — |

[a] Model Eq. (6) including spectral diffusion.
[b] Model Eq. (7) without spectral diffusion.

The results are reported in Table I, examples of fitted curves can be found in Figure 1 and the variation of $T_m$ as a function of $\beta$ is reported in Figure 2. It can easily be seen that instantaneous diffusion has a strong impact on the observed echo decays. For $\beta = \pi$, the exponential decay time $T_m$ is about 0.1 ms, irrespective of the isotopic composition, and sets a strong limit to quantum computing operations. At the limit $\beta \to 0$, the intrinsic phase memory time $T_{m0}$ is considerably longer, shows a more pronounced variation with temperature and isotopic composition, and even approaches the spectral diffusion time $T_{SD}$. The latter is longer in Si-28 than in Si-nat, the ratio $T_{SD}$(Si-28)/$T_{SD}$(Si-nat) being 1.4 and 2.5 at 10 K and 7 K, respectively. A recent theory[22] of nuclear spin induced spectral diffusion of phosphorus donor in silicon however estimates $T_{SD}$(Si-28)/$T_{SD}$(Si-nat) ≈ 100 when [$^{29}$Si] decreases from the natural abundance 4.67% to less than 0.1% in the enriched sample. The estimated ratio is then much larger than that observed. Moreover, $T_{SD}$ is independent of temperature in Si-nat within experimental errors, but not in Si-28. These findings and the Lorentzian shape of the field-swept ESE spectrum of Si-28[7] prompted us to check if spectral diffusion is really needed for a full description of the echo decays in Si-28. Therefore, a fit was carried out assuming an infinitely long $T_{SD}$, i. e., with a purely exponential model

$$I_e(\tau,\beta,T) = I_e(0,\beta,T) \exp\left\{-2\tau\left[\frac{1}{T_{m0}(T)} + C_1 \sin^2\left(\frac{\beta}{2}\right)\right]\right\} \quad (7)$$

The results are reported in Table I and the corresponding curves in Figure 1c,d where they exactly overlap with those obtained by Eq. (6). Of course, the values of $T_{m0}$ are shorter in this case because of the reduced number of optimized parameters. Therefore, even if there is no direct evidence that spectral diffusion is ineffective in relaxing the electron spins in the Si-28 sample, the experimental data can be adequately described without resorting to spectral diffusion, at variance with Si-nat.

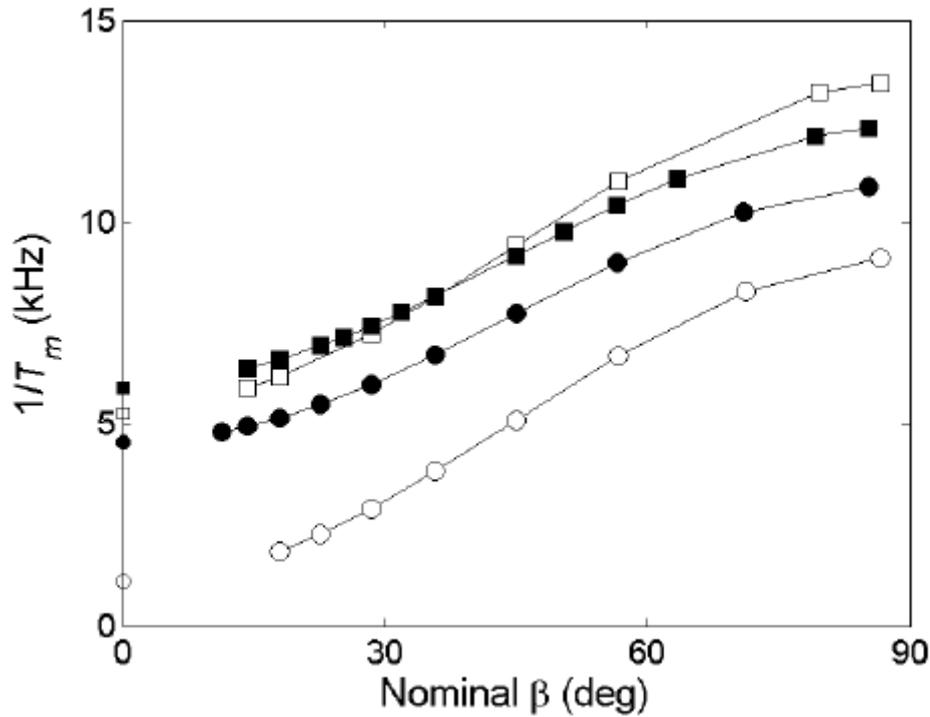

**FIG 2.** Effect of the nominal (on-resonance) second pulse turning angle β on the phase memory time $T_m$ of Si-nat and Si-28 samples observed by the β/2–τ–β–τ–echo pulse sequence. Open circles: Si-28 at 7 K; open squares: Si-28 at 10 K; solid circles: Si-nat at 7 K; solid squares: Si-nat at 10 K. The intrinsic phase memory time $T_{m0}$ is shown as smaller symbols on the vertical axis. The $T_m$ values are computed from the fitted slope factor $C$ and the $T_{m0}$ values obtained from optimization procedures including spectral diffusion in the fit model.

The phosphorus concentration obtained by the above fitting procedure is in all cases lower than the one measured by the Hall effect. This is probably due to the assumption underlying the derivation of Eqs. (3) and (4), i. e., that the phosphorus atoms are homogenously distributed over the entire sample volume. Of course, phosphorus atoms can only be found at the lattice sites of the silicon crystal, but such a discrete distribution does not seem to be relevant for the explanation of the observed discrepancy.[26] Much more important is the fact that the assumption allows two phosphorus atoms to be close to each other. This can occur in our samples, but such close pairs do not resonate in the same experimental conditions as isolated electron spins do and, therefore, do not contribute to the observed instantaneous diffusion. Hence, Eqs. (3) and (4) take into account very short spin-spin distances which greatly enhance the effectiveness of instantaneous diffusion. This results in an underestimation of the actual phosphorus concentration.

Finally, a comparison of the present results with literature data (See Table II) may be useful even if the experimental conditions (phosphorus concentration and temperature) are different. As for natural silicon samples, our $T_m$ and $T_{SD}$ values are in line with literature data. Note that the insensitivity of the data to temperature and [P] is due to the prevalence of instantaneous diffusion as relaxation mechanism. The close relationship between $T_{SD}$ and the isotopic composition of silicon is further supported by the investigation[16] of a $^{29}$Si-enriched sample with $T_{SD} = 0.019$ ms. Our data related to $^{28}$Si-enriched samples are in line with those presented in the pioneering paper by Gordon and Bowers[9] but disagree with recent data published by Tyryshkin et al.[15] The discrepancy can only arise from a difference in the donor concentration, which is about 40 times larger in the presently studied samples. This would imply that $T_{m0}$ values are sensitive to the donor concentration when magnetic nuclei are rare. Since in our sample the estimated average P-P distance is significantly larger than the effective Bohr radius of P in Si (2.1 nm), such a dependence can not arise from the overlap of the donor wavefunctions, but it is rather caused by dipolar coupling. In Ref. 15, the phosphorus concentration determined from instantaneous diffusion data agrees with the one obtained from Hall-effect measurements. In the last case, the average P-P distance is 3.5 times larger than in our case, so that the above described effect due to the "magnetically excluded volume" is less important.

**Table II**. Literature electron spin relaxation data for phosphorus shallow donors in silicon, measured by electron paramagnetic resonance methods.

| Sample | $T$ (K) | $T_m$ (ms) | [P] = 2 $C$ ($10^{16}$ cm$^{-3}$) | [P] from Hall effect ($10^{16}$ cm$^{-3}$) | $T_{SD}$ (ms) | Ref. |
|---|---|---|---|---|---|---|
| Natural Si | 8 | | | 0.18 | 0.25 [a] | 16 |
| | 7-12 | | | 0.08 | 0.63 | **15** |
| | 1.6 | 0.30 | | 1 – 6.6 | 0.18 | 11 |
| | 1.4 | 0.24 | | 3 | | 10 |
| | 1.4 | 0.20 | | 10 | | 10 |
| $^{28}$Si enriched Si | 8.1 | 14 ± 2 [b] | 0.087± 0.05 | 0.08 | | **15** |
| | 6.9 | 62 (+50/–20) [b] | 0.087± 0.05 | 0.08 | | **15** |
| | 1.4 | 0.52 | | 4 | | 10 |
| $^{29}$Si enriched Si | 8 | | | 0.18 | 0.019 [a] | 16 |

[a] $\exp(-b\tau^2)$ decay
[b] $T_{m0}$ value

## 3.2 Electron spin echo modulation

Fourier transformation of the electron spin echo envelope modulation (ESEEM) yields a spectrum that displays the transition frequencies $\nu_\alpha$ and $\nu_\beta$ of nuclear spins coupled to the unpaired electron spin:

$$\left.\begin{array}{c}\nu_\alpha \\ \nu_\beta\end{array}\right\} = \sqrt{\left(\nu_I \pm \frac{A}{2}\right)^2 + \left(\frac{B}{2}\right)^2} \qquad (8)$$

where $A$ and $B$ depend on the elements of the hyperfine interaction matrix and $\nu_I$ is the nuclear Larmor frequency. Peaks at the sum and difference of these transition frequencies ($\nu_\pm = \nu_\alpha \pm \nu_\beta$) are also observed in the two-pulse ESEEM spectrum.[23] Preliminary results[7] have shown that the ESEEM of the Si-nat sample is caused by the interaction between the donor electron spins and the $^{29}$Si nuclear spins. We here present a thorough analysis of the ESEEM spectrum of Si-nat, including accurate spectral fitting based on the hyperfine interaction matrices measured and assigned to specific crystal sites by ENDOR spectroscopy.[9,14,19] In the following, the nomenclature used in these papers is employed. In particular we refer to shells of crystallographically equivalent lattice sites about the donor impurity as: A shell (class <001>), comprising sites at (0,0,4) and symmetry-related positions; B shell (class {110}), comprising sites at (4,4,0) and symmetry-related positions, and E shell (class <111>), comprising sites at (1,1,1) and symmetry-related positions.

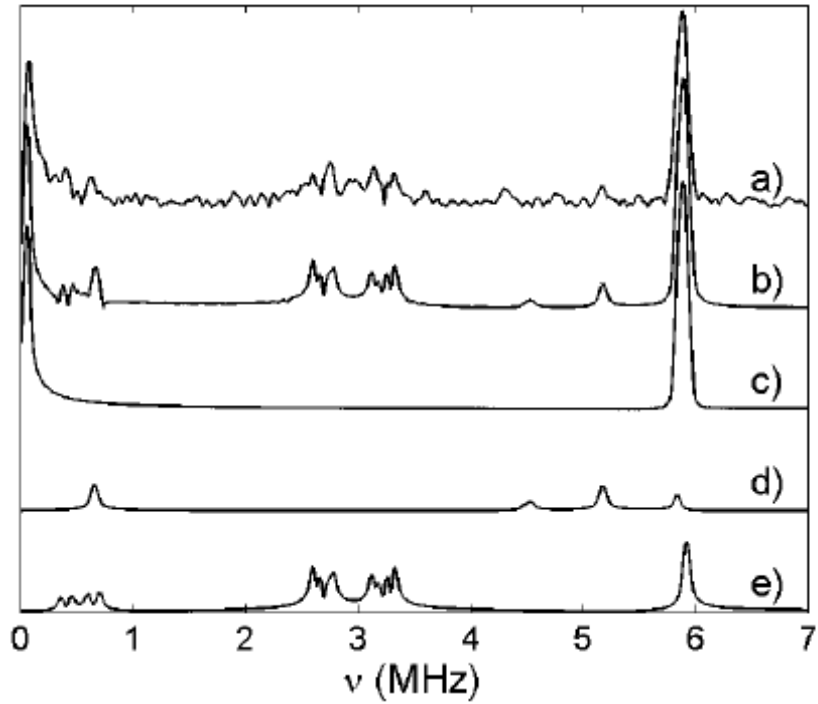

**FIG 3.** Two-pulse ESEEM spectrum of Si-nat at 10 K. All spectra are apodized by a Kaiser $2\pi$ window and zero-filled up to 4096 points a) Experimental spectrum; b-e) results of spectral fitting (time traces have been truncated in order to reproduce deadtime effects as in the experimental spectrum): b) total spectrum, c) contribution from A shell; d) contribution from B shell; e) contribution from E shell.

The ESEEM spectrum of Si-nat, reported in Figure 3, features several pairs of lines, most symmetrically placed about the Larmor frequency of $^{29}$Si (2.92 MHz). Thanks to the relatively low number of lines, some of them can tentatively be assigned to specific sites by comparison with the ESEEM frequencies reported in Table III. The lines between 2.5 and 3.5 MHz should arise from the E shell, except for the innermost partly resolved pair. The lines with frequencies 0.7 and 5.7 MHz might come from the B shell. The intense line close to zero frequency could be assigned to the $\nu_\alpha$ transition of the A shell, since the $^{29}$Si in those positions have hyperfine interactions matching two times the $^{29}$Si nuclear Zeeman interaction, thus causing a large modulation depth. The other intense line close to twice the $^{29}$Si Larmor frequency cannot even tentatively assigned since many ESEEM lines are predicted to lie in that range (see Table III), namely the $\nu_\beta$, $\nu_+$, and $\nu_-$ lines of the A shell and the $\nu_+$ lines of all the sites with significant ESEEM effect.

**Table III**. ESEEM frequencies ν (MHz) and modulation depth parameter $k$ for $^{29}$Si in Si:P with orientation (θ = 5.6°, ϕ = 251.6°) and static magnetic field corresponding to ν($^{29}$Si) = 2.92 MHz.

| Shell, Class | Site | $\nu_\alpha$ | $\nu_\beta$ | $\nu_+$ | $\nu_-$ | $k$ |
|---|---|---|---|---|---|---|
| A, <001> | 1 | 0.08 | 5.92 | 6.00 | 5.84 | 0.0009 |
|  | 2 | 0.05 | 5.89 | 5.94 | 5.84 | 0.1543 |
|  | 3 | 0.05 | 5.89 | 5.94 | 5.84 | 0.1448 |
|  | 4 | 0.08 | 5.92 | 6.00 | 5.84 | 0.0030 |
|  | 5 | 0.06 | 5.89 | 5.95 | 5.84 | 0.1187 |
|  | 6 | 0.06 | 5.89 | 5.95 | 5.84 | 0.1325 |
| B, {110} | 1 | 0.68 | 5.16 | 5.84 | 4.48 | 0.0033 |
|  | 2 | 0.68 | 5.16 | 5.84 | 4.48 | 0.0027 |
|  | 3 | 0.69 | 5.15 | 5.84 | 4.47 | 0.0017 |
|  | 4 | 0.69 | 5.15 | 5.84 | 4.47 | 0.0012 |
|  | 5 | 0.67 | 5.17 | 5.84 | 4.51 | 0.0088 |
|  | 6 | 0.66 | 5.18 | 5.84 | 4.52 | 0.0096 |
|  | 7 | 0.65 | 5.19 | 5.84 | 4.54 | 0.0095 |
|  | 8 | 0.66 | 5.18 | 5.84 | 4.52 | 0.0091 |
|  | 9 | 0.67 | 5.17 | 5.84 | 4.50 | 0.0085 |
|  | 10 | 0.65 | 5.19 | 5.84 | 4.54 | 0.0098 |
|  | 11 | 0.65 | 5.19 | 5.84 | 4.55 | 0.0096 |
|  | 12 | 0.67 | 5.18 | 5.84 | 4.51 | 0.0087 |
| E, <111> | 1 | 2.77 | 3.14 | 5.91 | 0.37 | 0.0939 |
|  | 2 | 2.74 | 3.18 | 5.92 | 0.44 | 0.1022 |
|  | 3 | 2.66 | 3.27 | 5.93 | 0.61 | 0.1163 |
|  | 4 | 2.61 | 3.32 | 5.93 | 0.71 | 0.1215 |

These tentative assignments can be confirmed and the nature of other lines understood only by careful spectral fitting of the entire spectrum. Fitting of ESEEM spectra is based on the hyperfine matrices[9,14,19] and the analytical formulae[23] given in the literature. The fitting parameters are the polar angles θ and ϕ of the static magnetic field with respect to the crystal reference frame, along with trivial offset and scale factors. Because of the dead-time in the ESEEM experiment, one has to resort to magnitude presentation of the ESEEM spectra. This may cause erroneous spectral interpretation due to distortions and artifacts.[27] As detailed in the Experimental section, the ESEEM time trace is first computed and, before and during the Fourier transformation, subject to the same manipulations as the experimental data, including truncation of the initial part of the computed trace. Hence, the fitting *results* are not affected by deadtime effects since artifactual distortions are part of the model itself.

The results of fitting the experimental spectrum (apodized by a Kaiser 2π window) in the frequency range 0-7 MHz is shown in Figure 3; the optimized angles are θ = (5.6±0.3)° and ϕ = (252±1)°. The match of experimental and computed spectra is very good for both line positions and intensities. The decomposition of the ESEEM spectrum into site-specific contributions (see Figure 3) led to the following assignments: line at 0.1 MHz, $\nu_\alpha(A)$; multiplet in the 0.3-0.7 MHz range, $\nu_-(E)$ and $\nu_\alpha(B)$; multiplets at about 2.8 and 3.2 MHz, $\nu_\alpha(E)$ and $\nu_\beta(E)$, respectively; line at 5.2 MHz, $\nu_\beta(B)$. The broad line at 5.9 MHz is an unresolved multiplet comprising the $\nu_\beta(A)$, $\nu_-(A)$ and $\nu_+$(all shells) lines.

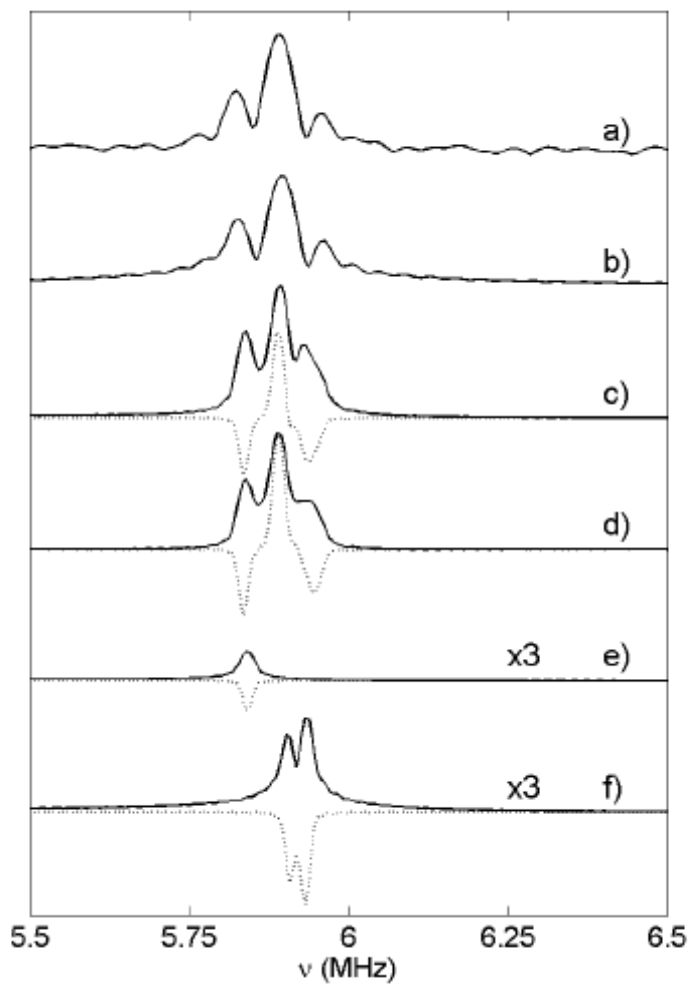

**FIG. 4.** Two-pulse ESEEM spectrum of Si-nat at 10 K. Solid lines: magnitude mode, dotted lines: absorption mode. All spectra are resolution enhanced by a growing exponential (time constant 15 µs) and zero-filled up to 4096 points. a) Experimental spectrum; b) Fitted spectrum (deadtime truncated signal). c-f) Ideal spectra (no deadtime effect): c) total spectrum, d) contribution from A sites; e) contribution from B sites; f) contribution from E sites. Spectra e) and f) are multiplied by 3 for the sake of better presentation.

To improve resolution in the 5-7 MHz spectral region, the experimental time trace was multiplied by a growing exponential prior to FT, thus trading sensitivity for resolution. This presentation of the experimental data is shown in Figure 4a, where three lines can be easily discerned. The results of fitting this resolution-enhanced spectrum are reported in the same Figure. Although a simulation carried out with the previously optimized angles already gave a good agreement with the experimental spectrum, the optimization procedure was repeated and yielded a very good agreement with the experimental data with optimized angles $\theta = (8.0\pm0.8)°$ and $\phi = (265\pm5)°$. In order of increasing frequency, the three resolved features are dominated by the $\nu_-$, $\nu_\beta$, and $\nu_+$ lines of the A shell, respectively. Besides, there are significant contributions from the $\nu_+(B)$ and $\nu_+(E)$ transitions. Note also that in this case the peak maxima do not correspond to ESEEM frequencies and deadtime distortions are important. Hence, correct assignment is only possible by accurate spectral fitting. For instance, the trough between the second and third feature is due to absorption/dispersion mixing caused by the deadtime truncation since the non-truncated signal has a maximum there in both cosine and magnitude mode. There is a slight difference between the optimum angles obtained in the two fitting procedures that can be traced back to (i) the small anisotropy of the hyperfine interactions and (ii) the presence of many chemically equivalent but magnetically nonequivalent sites for each shell, which make line positions and intensities rather insensitive to the orientation of the static field. However, the close similarity of the optimized angles from the two procedures allows us to be confident in the fitting model and results.

## 4 Conclusions

In this paper the P shallow donors in natural and in isotopically purified silicon have been investigated using pulse electron paramagnetic resonance spectroscopy. Electron spin phase relaxation times have been measured in natural and $^{28}$Si enriched silicon epilayers doped with $10^{16}$ cm$^{-3}$ P donors. Two-pulse ESE experiments performed as a function of the pulse turning angle allowed us to measure the exponential and spectral-diffusion relaxation times depurated by the additional effects of instantaneous diffusion. Spectral diffusion is an important dephasing mechanism which is significantly reduced in the isotopically purified Si-28 sample. The influence of the second pulse, in any sequence aiming at qubit manipulation, must be taken into consideration as, depending also on the total P concentration, instantaneous diffusion could reduce dramatically the phase coherence time. According to these results, isotopically purified samples are necessary to reduce the spectral diffusion contribution and, from a comparison with the literature data, the P shallow donors concentration plays a fundamental role to determine the intrinsic phase memory time of these materials.

Accurate analysis of the ESEEM spectra, observed in the Si-nat natural silicon sample, provides information on the shallow donor wave function. ESEEM peaks have been attributed to the hyperfine-coupled silicon-29 nuclei in the various crystallographic positions on the basis of a spectral fit procedure including instrumental distortions.

## References

[1]B. E. Kane, Nature (London) 393, 133 (1998)
[2]R. Vrijen, E. Yablonovitch, K. Wang, H. W. Jiang, A. Balandin, V. Roychowdhury, T. Mor, D. DiVincenzo, Phys. Rev. A 62, 012306 (2000)
[3]B. E. Kane, Fortschr. Phys. 48, 1023 (2000)
[4]B. E. Kane, N. S. McAlpine, A. S. Dzurak, and R. G. Clark, G. J. Milburn, He Bi Sun, H. Wiseman, Phys. Rev. B 61, 2961 (2000)
[5]M. Friesen, P. Rugheimer, D. E. Savage, M. G. Lagally, D. W. van der Weide, R. Joynt, M.A. Eriksson, Phys. Rev. B 67, 121301 (2003)

[6] J. Preskill, Proc. R. Soc. London, Ser. A 454, 385 (1998)

[7] M. Fanciulli, P. Höfer, A. Ponti, Physica B 340, 895 (2003)

[8] E. Yablonovitch, H. W. Jiang, H. Kosaka, H. D. Robinson, D. S. Rao, T Szkopek, Proc. IEEE, 91, 761 (2003)

[9] G. Feher, Phys. Rev. 114, 1219 (1959)

[10] J. P. Gordon and K.D. Bowers, Phys. Rev. Lett. 15, 368 (1958)

[11] M. Chiba and A. Hirai, J. Phys. Soc. Japan 33, 730 (1972)

[12] J. R. Klauder and P.W. Anderson, Phys. Rev. 125, 912 (1962)

[13] S. Saikin and L. Fedichkin, Phys. Rev. B 67, 161302(R) (2003)

[14] E. B. Hale and R. L. Mieher, Phys. Rev. 184, 739 (1969)

[15] A. M. Tyryshkin, S. A. Lyon, A. V. Astashkin, A. M. Raitsimring, Phys. Rev. B 68, 193207 (2003)

[16] E. Abe, K. M. Itoh, J. Isoya, S. Yamasaki, Phys. Rev. B 70, 033204 (2004)

[17] F. F. Kuo and J. F. Kaiser (Eds.), System Analysis by Digital Computer, Wiley, New York, USA (1967).

[18] R. R. Ernst, G. Bodenhausen, A. Wokaun, Principles of Nuclear Magnetic Resonance in One and Two Dimensions, Clarendon Press, Oxford (1991), p. 106.

[19] J. L. Ivey and R. L. Mieher, Phys. Rev. B 11, 849 (1975)

[20] J. A. Nelder and R. Mead, Computer J. 7, 308 (1965); J. C. Lagarias, J. A. Reeds, M. H. Wright and P. E. Wright, SIAM J. Optim. 9, 112 (1998).

[21] W. H. Press, B. P. Flannery, S. A. Teukolsky, W. T. Vettering, Numerical recipes in C, Cambridge University Press, Cambridge, UK (1988).

[22] R. de Sousa and S. Das Sarma, Phys. Rev. B 68, 115322 (2003)

[23] A. Schweiger and G. Jeschke, Principles of Pulse Electron Paramagnetic Resonance, OUP, Oxford, UK (2001)

[24] A. M. Raitsmiring, K.M. Salikhov, B.A. Umanskii, Y.D. Tsetkov, Sov. Phys. Solid State 16, 492 (1974)

[25] W. A. Coish and D. Loss, Phys. Rev. B 70, 195340 (2004)

[26] W. B. Mims in Electron Spin Resonance (Ed.: S. Geschwind), Plenum Press, New York, USA (1972)

[27] S. Van Doorslaer, G. A. Sierra, A. Schweiger, J. Magn. Reson. 136, 152 (1999)